\documentclass[pre,aps,twocolumn]{revtex4}
\usepackage{epsfig}
\usepackage{bm}
\newcommand{\B}[1]{{\bm{#1}}}

\usepackage[latin1]{inputenc}
\newcommand{\beq}{\begin{equation}}
\newcommand{\eeq}{\end{equation}}
\newcommand{\bea}{\begin{eqnarray}}
\newcommand{\eea}{\end{eqnarray}}
\begin{document}
\title{Branching Instabilities in Rapid Fracture: Dynamics and
Geometry}
\author{Eran Bouchbinder, Joachim Mathiesen and Itamar Procaccia}
\affiliation{Dept. of Chemical Physics, The Weizmann Institute of
Science, Rehovot 76100}
\begin{abstract}
We propose a theoretical model for branching instabilities in
2-dimensional fracture, offering predictions for when crack
branching occurs, how multiple cracks develop, and what is the
geometry of multiple branches. The model is based on equations of
motion for crack tips which depend only on the time dependent stress
intensity factors. The latter are obtained by invoking an
approximate relation between static and dynamic stress intensity
factors, together with an essentially exact calculation of the
static ones. The results of this model are in good agreement with a
sizeable quantity of experimental data.
\end{abstract}
\maketitle

\section{Introduction}

The phenomenon of crack division, i.e. the splitting of a single
primary crack into two or more branches, whose dynamics develops
independently, is studied in thin plates of different materials
(glasses, plastics, metals etc.) \cite{73Kal}. A classical example
from more than thirty five years ago is shown in Fig. 1, which
exhibits a crack pattern observed in a thin araldite plate
\cite{69And}.
%%%%%%% FIGURE 1 %%%%%%%%%%%%%%%%%%
\begin{figure}
\centering \epsfig{width=.45\textwidth,file=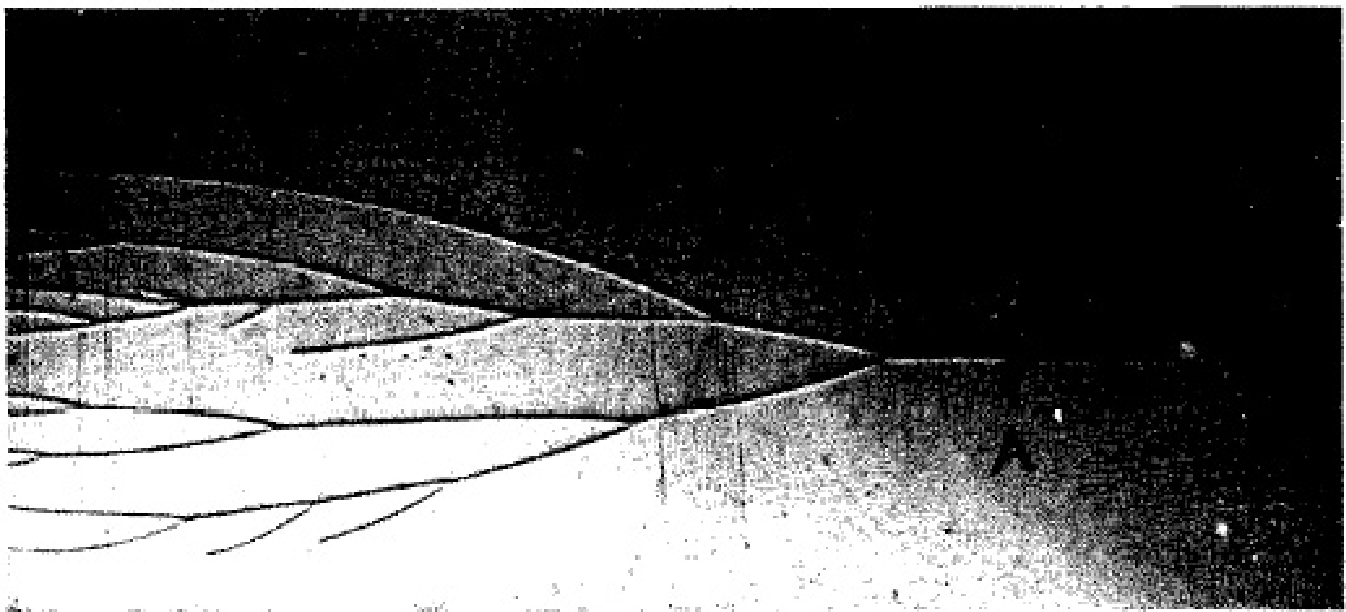}
\caption{Crack pattern in an araldite tensile sheet \cite{69And}. }
\label{Andersson}
\end{figure}
%%%%%%%%%%%%%%%%%%%%%%%%%%%%%%%%%%
In this example the cracks go throughout the thickness of the sheet.
This is to be distinguished from apparently similar side branching
instabilities in which side branches appear in addition to the main
crack, when the velocity of propagation exceeds a critical value
\cite{95SGF,96SF}. There exists an important difference between the
two phenomena: in the former cases all the branches crack through
the plates (some time referred to as ``macrobranching"), whereas in
the latter experiments, near the onset of the instability, the side
branches have a width which is considerably smaller than the thickness
of the plate (some time referred to as ``microbranching").
Notwithstanding attempts to interpret the latter phenomenon using
2-dimensional theories \cite{04ABa, 04ABb}, it appears that the
interaction of crack fronts of different thickness necessitates a
3-dimensional theory which is daunting at present. In this paper we
limit our discussion to a 2-dimensional theory that pertains only to
plates with cracks going through the plate.

The aim of this paper is to develop an approximate theory of crack
dynamics, including crack bifurcations and multiple crack
competition. More specifically, we address the following questions:
\begin{enumerate}
\item Given a straight propagating crack in a 2-dimensional material,
when and how the first bifurcation occurs? The bifurcation event
itself was studied successfully in a recent paper by Adda-Bedia
\cite{04ABa}. In our approach we are able to examine the dynamics of
the bifurcated cracks.

\item Given a bifurcated crack, what is the stability of a symmetric
branched configuration?
We show that there exists an instability towards geometric
perturbations, making one branch growing on the expense of the other
which gets arrested.

\item What is the geometry of multiple (two or more) branches?
Obviously the interaction between multiple
branches results in curved cracks. The model that we propose is able
to follow the dynamics and the resulting geometry.

\item How consecutive bifurcations come about, and what are the
resulting crack patterns?

\item Can one apply the 2-dimensional theory developed here to
microbranching? The answer
will be shown to be negative; there are crucial 3-dimensional
aspects of microbranching that need the 3-dimensional theory in its
full galore for an appropriate treatment.
\end{enumerate}

In order to answer all these questions, we take the point of view
that the dynamics of the tip of each crack is determined by the
elastic field in its very vicinity. Near each tip one expands the
stress field as usual,
\begin{equation}
\sigma_{ij}(r,\theta,t) = K_{_{\rm I}}(t) \frac{\Sigma^{^{\rm
I}}_{ij}(\theta,v)}{\sqrt{2\pi r}}+K_{_{\rm II}}(t)
\frac{\Sigma^{^{\rm II}}_{ij}(\theta,v)}{\sqrt{2\pi r}} \ .
\label{sigma}
\end{equation}
Here $v$ is the instantaneous tip velocity, $\{r, \theta\}$ are
polar coordinates at the crack tip and $t$ is time. $ K_{_{\rm
I}}(t)$ and $ K_{_{\rm II}}(t)$ are the {\em dynamic} stress
intensity factors, and the functions $\B \Sigma$ are known universal
functions of $\theta$ and $v$.

The central element in our model is the adoption of the
Hodgdon-Sethna equations for the crack tips \cite{93HS}. These
equations were derived in the context of quasi-static crack
propagation, and were shown to be in agreement with quasi-static
experiments in \cite{03BHP}. Here we employ these equations in the
dynamic context, invoking the results and their comparison with
experiments for justification. Consider a local coordinates system
located at the crack tip, in which $\hat{\bf t}$ and $\hat{\bf n}$
denote the tangential and normal directions respectively. For the
crack tip location ${\bf r}^{tip}$ we write the equations
\begin{eqnarray}
\label{HS}
\frac{\partial {\bf r}^{tip}}{\partial t} &=&  v \hat{\bf t} \nonumber
\\
\frac{\partial \hat{\bf t}}{\partial t}  &=&  - f K_{_{\rm II}}(t)
\hat{\bf n} \ .
\end{eqnarray}
Here $v$ is the instantaneous crack tip velocity, and $ f$ is a
positive material parameter. Needless to say, these equations appear
simpler than they really are. The actual calculation of the dynamic
stress intensity factors for evolving cracks of complicated geometry
is very far from trivial. The bulk of Sect. \ref{dymcrack} is devoted
to the presentation of approximate schemes to compute these objects.
The main idea is to relate the dynamic stress intensity factors to
their static counterpart \cite{04BKP}, and then to compute the
latter using the method of iterated conformal maps that had been
presented for fracture problems in recent papers
\cite{04BMPa,04BMPb}. This material is reviewed in Sect. \ref{stat}.

Having at hand the stress intensity factors one can use them in Eqs.
(\ref{HS}) for each crack tip. In Sect. \ref{moredym} we describe
how this is done for multiple branch dynamics. We demonstrate that
the theory is successful in describing crack bifurcations, crack
arrest, and successive bifurcations. The geometry can be studied in
great detail and compared (very favorably) with available
experiments. Finally, in Sect. \ref{micro} we consider the
applicability of this theory to microbranching. As mentioned above,
the conclusion is negative. Sect. \ref{summary} offers a summary of
the paper and some concluding remarks.
%%%%%%%%%%%%%%%%%%%%%%%%%%%%%%%%%%%%%%%%%%%%%%%%%%%%
\section{Dynamics of multiple cracks}
\label{dymcrack}
To implement Eqs. (\ref{HS}) we need first to compute the dynamic
stress intensity factors, and second the velocity $v$ at the tip of
each branch in a multiple branch configuration. We start with the
stress intensity factors.
%%%%%%%%%%%%%%%%%%%%%%%%%%%%%%%%%%%%%%%%%%%%%%%%%%%%%%%%%%%%%%%
\subsection{Estimating the Dynamic Stress Intensity Factors}
\label{dym} The first task is the calculation of the {\em dynamic}
stress intensity factors for a branched configuration. To this aim
we invoke the formalism developed in \cite{04BKP}, in which the
dynamic stress intensity factors were related to their static
counterparts. Based on some specific examples an approximate form of
the {\em dynamic} stress intensity factors was obtained as a product
of the {\em static} stress intensity factors $K^s_{_{\rm I}}$ and
$K^s_{_{\rm II}}$ (of the instantaneous frozen configuration) and
universal functions of the instantaneous velocity. For each mode of
fracture one writes
\begin{eqnarray}
K_{_{\rm I}}(t) &\simeq& k_{_{\rm I}}(v) K^s_{_{\rm I}}(t) \nonumber\\
K_{_{\rm II}}(t) &\simeq& k_{_{\rm II}}(v) K^s_{_{\rm II}}(t) \ .
\label{decopm}
\end{eqnarray}
We note that both the dynamic and the static stress intensity
factors are considered time-dependent. For the static objects this
dependence means freezing the actual configuration that is obtained
at time $t$.  The universal functions of the velocity are given by
\begin{eqnarray}
k_{_{\rm I}}(v) &=&
S\left(-\frac{1}{v}\right)\frac{1-v/c_{R}}{\sqrt{1-v/c_{d}}}
\nonumber\\
k_{_{\rm II}}(v) &=&
S\left(-\frac{1}{v}\right)\frac{1-v/c_{R}}{\sqrt{1-v/c_{s}}}
\end{eqnarray}
and $S(\zeta)$ is given by
\begin{widetext}
\begin{equation}
S(\zeta) = \exp\left[-\frac{1}{\pi}\int_{1/c_{d}}^{1/c_{s}}\tan^{-1}
\left(\frac{4\eta^2\sqrt{\left(\eta^2-c^{-2}_d\right)\left(c^{-2}_{s}-\eta^2\right)}}
{\left(c^{-2}_{s}-2\eta^2\right)^2}\right)
\frac{d\eta}{\zeta+\eta}\right].
\end{equation}
\end{widetext}
Here $c_R, c_{d}$ and $c_{s}$ are the Rayleigh, dilatational and
shear wave speeds respectively.

This approximation was shown in the classical theory of fracture
mechanics to be essentially exact for semi-infinite straight cracks
\cite{98Fre}. In our earlier work we applied this approximate
methodology for describing interacting large and small cracks
\cite{04BKP}. The separable form of the dynamic stress intensity
factors is trivially correct for very small velocities, since the
functions $k_I$ and $k_{II}$ tend to unity for $v\to 0$. For finite
velocities the separable form is not exact, but we expect it to
yield good approximations when the velocities are small fractions of
the typical wave speed and the typical distance between the evolving
tips is small. We expect this since the mutual information about the
location of each tip is carried by waves; under the specified
conditions these waves can deliver the required mechanical
information before a substantial change in the state of the system
took place. We will show below that these conditions are satisfied
in the early stages of the branching instability, therefore allowing
us to use the separable form quite confidently in the present
context. It is difficult to quantify {\em a-priori} the range of
validity of the approximation. Therefore we invoke the final results
and their agreement with various experimental results to support the
quality of the approximation.

Clearly, this approximation is a huge simplification, calling for
solving the static equilibrium field equations rather than the full
dynamical field equations.  Of course,  one still has to face the
difficult problem of {\em static} non-trivial geometries, but this
problem was solved quite {\em generally} using the method of
iterated conformal maps \cite{04BMPa}, and demonstrated
in the context of complex crack geometries in \cite{04BMPb}.
%%%%%%%%%%%%%%%%%%%%%%%%%%%%%%%%%%%%%%%%%%%%%%
\subsection{The velocity of the crack tips}

To close Eqs. (\ref{HS}) as a consistent mathematical system we need
to compute the velocity of each tip in terms of the dynamic stress
intensity factors. The basic idea is to employ the energy balance
that equates the energy release rate into the crack tip region
(denoted by $G$) to the dissipation involved in the crack
propagation (denoted by $\Gamma$). The classical theory of linear
elasticity fracture mechanics \cite{98Fre} provides the energy
release rate into each tip region:
\begin{equation}
 G = \frac{1-\nu^2}{E}\left[ A_{_{\rm I}}(v) K^2_{_{\rm I}} +A_{_{\rm
II}}(v) K^2_{_{\rm II}}
 \right],\\
 \label{ERR}
\end{equation}
where $E$ and $\nu$ are the Young's modulus and Poisson's ratio
respectively and  $A_{_{\rm I}}(v)$ and $A_{_{\rm II}}(v)$ are
universal functions given by
\begin{eqnarray}
&& A_{_{\rm I}}(v) = \frac{v^2
\sqrt{1-v^2/c_{d}^2}}{(1-\nu)~c_{s}^2~D(v)} \ ,\label{Rayleigh}\\
&& A_{_{\rm II}}(v) = \frac{v^2
 \sqrt{1-v^2/c_{s}^2}}{(1-\nu)~c_{s}^2~D(v)} \ , \nonumber\\
&&D(v) =
4~\sqrt{\left(1-v^2/c_{d}^2\right)\left(1-v^2/c_{s}^2\right)}
-\left(2-v^2/c_{s}^2\right)^2 \ . \nonumber
\end{eqnarray}
Note that $D(v)$ vanishes at the Rayleigh wave speed, $v=\pm c_{R}$.

For concreteness, consider a two-dimensional infinite medium loaded
at infinity with a uniform constant tensile stress
$\sigma^{\infty}_{yy}$. A long straight crack propagates at an
instantaneous velocity $V$; at some critical velocity, when the
crack length is $L$, the crack bifurcates into two branches of
lengths $\{\ell_i\}\ll L$ with tip velocities $\{v_i\}$,  defining
angles $\{\lambda_i \pi\}$ with respect to the direction of the
crack prior to the bifurcation. Freezing the crack just at the
bifurcation we denote its static stress intensity factor as
$K^{(0)}_{_{\rm I}}$.

At each tip of the bifurcated crack we define the normalized stress
intensity factors
\begin{equation}
F_{_{\rm I}} = \frac{K^s_{_{\rm I}}}{K^{(0)}_{_{\rm I}}}\ , \quad
F_{_{\rm II}}  =  \frac{K^s_{_{\rm II}}}{K^{(0)}_{_{\rm I}}} \ .
\label{normSIF}
\end{equation}
Equating $G$ to $\Gamma$ we can rewrite Eq. (\ref{ERR}) at {\em each
branch tip} as
\begin{equation}
 \frac{E~\Gamma}{1-\nu^2} = g_{_{\rm I}}(v)F_{_{\rm
I}}^2\left[K^{(0)}_{_{\rm I}}\right]^2 +
 g_{_{\rm II}}(v)F_{_{\rm II}}^2\left[K^{(0)}_{_{\rm I}}\right]^2  \
, \label{Freund}
\end{equation}
with
\begin{eqnarray}
 g_{_{\rm I}}(v)&\equiv&A_{_{\rm I}}(v)k_{_{\rm I}}^2(v) \nonumber\\
 g_{_{\rm II}}(v)&\equiv&A_{_{\rm II}}(v) k_{_{\rm II}}^2(v) \ .
\end{eqnarray}
Under the assumption that $\Gamma$ is velocity independent we
observe that the left hand side of Eq. (\ref{Freund}) contains only
material parameters. Therefore, a similar equation holds for the
pure mode I crack propagating with velocity $V_b$ just before the
branching event,
\begin{equation}
 \frac{E~\Gamma}{1-\nu^2} =
 g_{_{\rm I}}(V_b)\left[K^{(0)}_{_{\rm I}}\right]^2.
\end{equation}
We conclude that the instantaneous velocity $v_i$ of each crack tip
is determined by $V_b$ according to
\begin{equation}
 g_{_{I}}(V_b) = g_{_{I}}(v_i)F_{_{I}}^2 + g_{_{II}}(v_i)F_{_{II}}^2 \
, \quad \forall i \ .
 \label{velocity}
\end{equation}
Note that it would be better to index the stress intensity factor
with an index $i$ to stress that each tip contributes its own
equation to the set. We avoid it in order not to overbear the
notation. Notwithstanding, in case for which the dissipation
function $\Gamma$ becomes velocity dependent, we should introduce
$\Gamma(v_i)/\Gamma(V_b)$ on the LHS of Eq. (\ref{velocity}).
Bearing in mind that $ g_{_{I}}(V_b)$ is a decreasing function, the
result of such a change would be a reduction in $V_b$ \cite{04ABa}.

On physical grounds one seeks solution of these relations for
non-negative branch velocities $v_i$. In ref. \cite{04ABa} it was
shown that under the {\em assumption} that the branches start their
evolution quasi-statically (i.e. $v_i=0$) a solution appears first
for symmetric branching at $V_b=v_c\approx 0.475 c_s$ . For higher
velocities $V_b>v_c$ one can have bifurcations in which the branches
start off at a finite velocity.

Finally, we bring the equations of motion to their final form as
used below. Let $\theta$ be the angle between the tangential unit
vector and the $x$-axis. Denoting the tip position of the straight
crack at bifurcation as $\B r_b$, $\ell_i \equiv |{\bf r}^{tip}_i
-\B r_b|$. We rewrite now the tangential and normal unit vectors at
the tip of the crack in terms $\theta$ and a re-scaled time $t
K^{(0)}_{_{\rm I}} f\to t$. These changes transform Eqs. (\ref{HS})
into
\begin{eqnarray}
 \label{Dynamics}
\frac{\partial \ell_i}{\partial t} &=&  v_i \nonumber \\
\frac{\partial\theta_i}{\partial t} &=& k_{_{II}}(v_i) F_{_{II}} \ .
\end{eqnarray}
These equations, in conjunction with Eqs. (\ref{velocity}), define
our dynamical system. Note that the velocities are measured in units
of $c_s$. We selected $\nu=0.25$ which implies $c_d=\sqrt{3}c_s$ and
$c_R=0.9194 c_s$. It is important to notice that the set of
equations for each branch tip is coupled to the equations for the other
tips via the functions $F_{_{I}}$ and $F_{_{II}}$. All the results
presented below are obtained by the following
procedure: for each instantaneous branched configuration the
functions $F_{_{I}}$ and $F_{_{II}}$ are calculated using the method
of iterated conformal mappings, then the velocities of the tips are
calculated using Eqs. (\ref{velocity}) and finally the increments in
length and angle are calculated according to Eqs. (\ref{Dynamics}).

\section{Static Branched Configurations}
\label{stat}

 A crucial ingredient in our model is the
calculation of the {\em static} stress intensity factors for an
arbitrary branched crack configuration. The general approach to this
problem, based on the method of iterated conformal mappings, is
presented in all detail in ref. \cite{04BMPa}. The essential
building block is the composition of the conformal map from the
exterior of the unit circle to the exterior of the complicated crack
shape, using a functional iteration of a fundamental conformal map
that adds one single bump to the unit circle. This scheme enables us
to solve for the entire stress field for any branched crack
configuration. The static stress intensity factors are extracted
from the near tip fields using the method explained in
\cite{04BMPa}. In the present context the crack
tips are dressed by a finite curvature determined by the size of the
bump of the fundamental map which is added at each
iteration. While appropriate for comparison with most realistic
experiments, where some blunting of the crack tip is always present,
this finite curvature means that comparison with mathematical models
with infinitely sharp cracks should be done with care.

\begin{figure}
\centering \epsfig{width=.45\textwidth,file=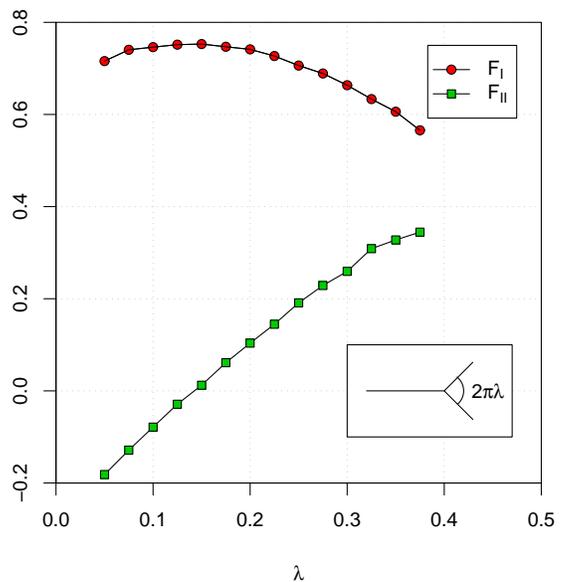}
\caption{The
normalized stress intensity factors $F_{_{I}}$ and $F_{_{II}}$ as a
function of $\lambda$, where $2  \pi \lambda$ is the angle between
the branches (see inset). The ratio of branch length and the main
crack is $\ell/L=0.5\times 10^{-3}$.}\label{symm}
\end{figure}

In the literature there is only limited amount of works presenting
calculations of stress intensity factors for in-plane problems with
branched crack configurations. In ref. \cite{04ABa} such a
calculation for an infinitesimal symmetric branched configuration is
provided using a numerical solution of an integral equation. In
order to ascertain the reliability and accuracy of our calculation
we consider first a similar configuration, choosing the length
$\ell$ of each branch such that $\ell/L=0.5\times 10^{-3}$. We could
not select a smaller ratio due to the finite curvature of the crack
tip (in addition, our numerical scheme which is based on truncated
Fourier expansions cannot deal efficiently with minute geometric
details, since the series truncation becomes inaccurate). The
normalized stress intensity factors $F_{_{I}}$ and $F_{_{II}}$ as a
function of $\lambda$, where $2  \pi \lambda$ is the angle between
the branches (see inset) are shown in Fig. \ref{symm}. This figure
should be compared with Fig. 4 in Ref. \cite{04ABa}. It is clear
that the figures are in good agreement (the locations of the maximum
of $F_{_{I}}$ and zero crossing of $F_{_{II}}$ are nearly
identical), although there is a slight overestimation in $F_{_{I}}$
due to finite size effects.
%%%%%% FIGURE 2 %%%%%%%%%%%%%%%%%%
%%%%%%%%%%%%%%%%%%%%%%%%%%%%%%%%%%%

\begin{figure}
\centering \epsfig{width=.45\textwidth,file=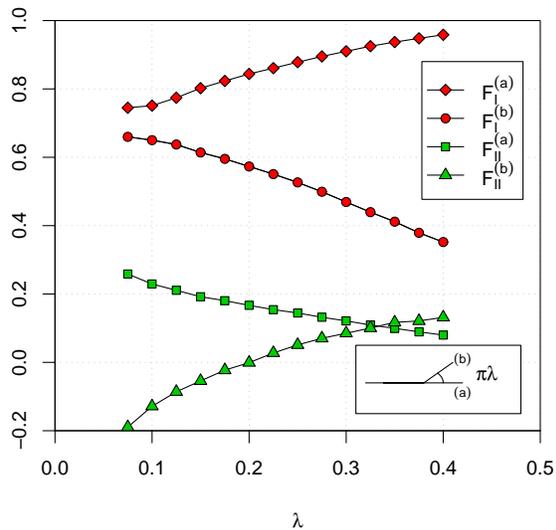}
\caption{The normalized stress intensity factors $F_{_{I}}$ and $F_{_{II}}$ for
both tips as a function of $\lambda$, where $\pi \lambda$ is the
angle between the branches (see inset). The ratio of branches length
and the main crack is $\ell/L=0.5\times 10^{-3}$.}\label{asymm}
\end{figure}

\begin{figure}
\centering \epsfig{width=.45\textwidth,file=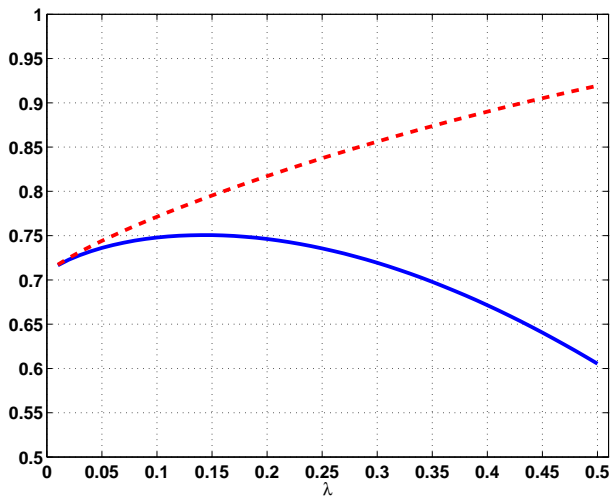}
\caption{The mode III normalized stress intensity factor for an
asymmetric branched configuration. $\pi \lambda$ is the angle
between the branches and the ratio of the branches length and the
main crack length is $\ell/L=0.5\times 10^{-3}$. The upper (lower)
curve corresponds to the forward
 direction (inclined) branch.}\label{AsymmSIFModeIII}
\end{figure}

Second, we considered an asymmetric branched configuration in which
both branches have the same length $\ell/L=0.5\times 10^{-3}$, while
one of
them is located in the direction of the main crack and the other
creates an angle of $\pi \lambda$ relative to that direction. The
normalized stress intensity factors $F_{_{I}}$ and $F_{_{II}}$ for
both tips as a function of $\lambda$ are shown in Fig. \ref{asymm}.
%%%%%%% FIGURE 3 %%%%%%%%%%%%%%%%%%
%%%%%%%%%%%%%%%%%%%%%%%%%%%%%%%%%%%
To our best knowledge there is no calculation available in the
literature for this configuration. Since mode III models are usually
in a qualitative agreement with their in-plane counterparts we
present in the appendix the calculation for a mode III asymmetric
branched configuration and present the resulting normalized stress
intensity factors in Fig. \ref{AsymmSIFModeIII}.
%%%%%%% FIGURE 4 %%%%%%%%%%%%%%%%%%
%%%%%%%%%%%%%%%%%%%%%%%%%%%%%%%%%%%
Indeed, the mode I component of the in-plane calculation shows the
same qualitative behavior as its mode III counterpart.
\section{The Dynamics of Multiple Branches}
\label{moredym}

In Ref. \cite{04ABa} it was found that under the conditions that the
branches start off quasi-statically (with zero velocity) and with
$K_{_{II}} \approx 0$, the critical velocity is $V_b=v_c \approx
0.475 c_s$ and the branching angle $\lambda\pi$ is $ 0.13\pi$. Note
that $\lambda$ is determined by the zero crossing of $K_{_{II}}$
presented in Fig. \ref{symm}. The velocity is ``critical" in the
sense that it is the first velocity for which branching is
energetically possible. Since there is no specification of the
mechanism of branching, one should treat it as a lower bound for the
branching velocity. For every $V_b>v_c$ one can find a solution with
the same $\lambda$ (since it is determined by the independent
condition $K_{_{II}} \approx 0$), but with a non-vanishing velocity of the branches.
Indeed, experimentally it appears that the branches do not
emerge quasi-statically as implied by the solution of \cite{04ABa}.
Note that as long as $V_b$ is not much larger than $v_c$, the
branches velocities are relatively small. Bearing in mind that the
distance between the tips is also relatively small, we can use the
separable form of the dynamic stress intensity factors quite
confidently. In this section we analyze the post-branching dynamics
for various physical conditions.

\subsection{Stability Analysis}

\begin{figure}
\centering \epsfig{width=.45\textwidth,file=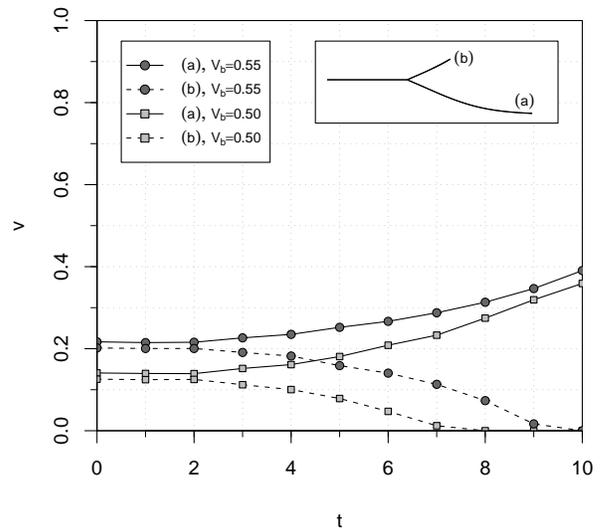}
\caption{The velocities of the branches as a function of time for $V_b=0.50$
and $V_b=0.55$. The resulting dynamics are such that the unperturbed
branch is arrested, while the perturbed one returns after a short
time to the original crack path. A representative resulting crack
pattern is shown in the inset. See text for details.}\label{instab}
\end{figure}

Motivated by the experimental evidence that symmetric branches do
not emerge quasi-statically, we study the stability of the symmetric
configuration against small geometrical perturbations. We consider
the possibility that dynamical instabilities prevent the development
of the branching event even though it is energetically allowed at
$V_b=v_c$. Consider a symmetric branched configuration with
$\ell/L=0.5\times 10^{-3}$ and $\lambda = 0.13$. Introduce a
positive small perturbation $\delta\ell=0.5\times 10^{-2}\ell$ to
the length of one of the branches, and integrate the dynamical Eqs.
(\ref{velocity}-\ref{Dynamics}) for various $V_b>v_c$. Note that as
$V_b$ increases so does the velocity of the emerging branches.
Figure \ref{instab} presents the resulting dynamics for $V_b=0.50$
and $V_b=0.55$. A representative resulting crack pattern is shown in
the inset; the unperturbed branch competes with the perturbed one
and eventually dies out (i.e. it gets arrested due to screening effects). The
velocities of the branches are plotted as a function of time. The
time to arrest can be identified as the point where the velocity of
the unperturbed branch vanishes. By comparing the data for the two
branching velocities, it is clear that the time to arrest increases
substantially as $V_b$ increases, therefore we deduce that the
increment of instability {\em decreases} with increasing $V_b$.
%%%%%%% FIGURE 5 %%%%%%%%%%%%%%%%%%
%%%%%%%%%%%%%%%%%%%%%%%%%%%%%%%%%%%
We conclude that symmetric branched configurations are unstable
against small geometrical perturbations at least for the branching
velocities we considered. We should stress that larger $V_b$'s are
not considered here due to the expected deterioration in the quality
of the approximation embodied in the separable form of the dynamic
stress intensity factors. In this regime it is reasonable to believe
that other dynamic effects are important and might stabilize the
symmetric configuration. The resulting dynamics are such that the
unperturbed branch is arrested, while the perturbed one returns
after a short time to the original crack path. Therefore, the main
effect of these attempted branching events is the sudden
deceleration of the crack. We suggest to interpret
the instability as a possible explanation for the fact that in
macrobranching events the branches do not emerge quasi-statically
since then the configuration is very sensitive to perturbations and
probably cannot be observed on macroscopic scales. On the other
hand, for larger branching velocities, for which the branches emerge
with finite velocities, both branches coexist, they are less
unstable to small perturbations and therefore can grow to observable
sizes.

\begin{figure}
\centering \epsfig{width=.45\textwidth,file=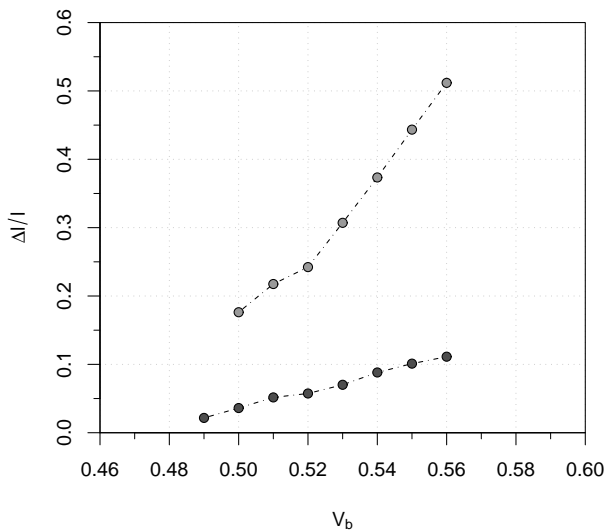}
\caption{The relative change in length $\Delta \ell/\ell$ as a
function of $V_b$ for $\delta\ell =2\times
10^{-2}\ell$ (lower curve) and $\delta\ell =0.5\times 10^{-2}\ell$
(upper curve). Note that the continuation of the lines intersects
the x-axis at the critical branching velocity $V_b=v_c \approx
0.475$, below which branching is energetically
forbidden.}\label{BranchLength}
\end{figure}

In the framework of stability analysis we also consider the final
length of the arrested branch. Denoting by $\Delta \ell$ the
difference between the final branch length and its initial length
$\ell$, we show in Fig. \ref{BranchLength} the relative change in
length $\Delta \ell/\ell$ as a function of $V_b$ for two values of
the fixed ratio $\delta\ell/\ell$. The dependence seems to be
approximately linear for both. Note that the continuation of the
lines intersects the x-axis at the critical branching velocity
$V_b=v_c \approx 0.475$, below which branching is energetically
forbidden. In passing, we note the resemblance of Fig.
\ref{BranchLength} to the experimental results obtained in
\cite{95SGF} for the microbranching instability. There it was found
(see Fig. 3 in \cite{95SGF}) that the branch length increases
approximately linearly with the mean crack velocity and vanishes for
the critical one.

%%%%%%% FIGURE 6 %%%%%%%%%%%%%%%%%%
%%%%%%%%%%%%%%%%%%%%%%%%%%%%%%%%%%%

%%%%%%%%%%%%%%%%%%%%%%%%%%%%%%%%%%%%%%%%%%%%%%%%%%%%%%%%%%%%%%%

\subsection{Successive Branching Events}

\begin{figure}
\centering \epsfig{width=.45\textwidth,file=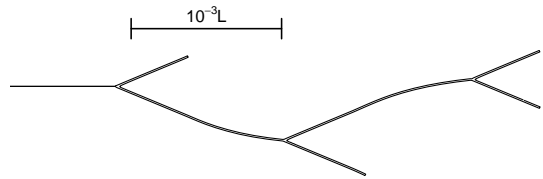}
\caption{The crack pattern that was formed by three successive
branching events. The branching velocity was set to $V_b=0.5$, i.e.
slightly more than $v_c$. We introduced a positive small
perturbation $\delta\ell=2\times 10^{-2}\ell$ to the length of the
lower branch. Upon reaching the branching velocity again we
introduced a symmetric configuration with the same $\ell$ and
$\lambda$ with no perturbation added by hand. The bar shows the scale
of the process relative to the initial crack length
$L$.}\label{Multiple}
\end{figure}

\begin{figure}
\centering \epsfig{width=.45\textwidth,file=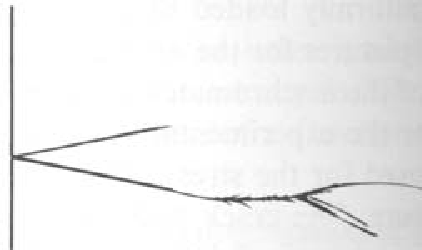}
\caption{A crack pattern in a tensile $260\times 120\times 5$ mm
araldite plate \cite{73Kal}. The plate was notched symmetrically at
the edge. Due to minute asymmetries in the production of the
notches, only one of them propagates. A second branching event can
be observed after some time. In this event almost symmetric branches
emerge from the branching point and coexist until one outruns the
other and then curves towards the symmetry line. Attempted branching
events can be clearly observed before the successful branching event
took place. All these features are in pleasing correspondence with
our discussion. }\label{Kalthoff1}
\end{figure}

In light of the observation of asymmetric branch growth with one of
them arrested, we follow now the evolution of the surviving branch.
This branch then accelerates to the critical branching velocity and
may bifurcate again. In this subsection we study the patterns formed
by such multiple successive branching events. In Fig. \ref{Multiple}
we present the crack pattern that was formed by three successive
branching events. At the first event we introduced a positive small
perturbation $\delta\ell=2\times 10^{-2}\ell$ to the length of the
lower branch. At the second event we introduced a symmetric
configuration with the same $\ell$ and $\lambda$ with no
perturbation added by hand. We stopped the evolution of the system at the onset of
the third event since within our numerical precision we could not
determine its outcome; in some cases it turned out that the upper
branch outruns the lower and in others vice versa.
%%%%%%% FIGURE 7 %%%%%%%%%%%%%%%%%%
%%%%%%%%%%%%%%%%%%%%%%%%%%%%%%%%%%%
The results shown in Fig. \ref{Multiple} are in qualitative
agreement with the experimental results shown in Fig.
\ref{Kalthoff1}. We note here that the seemingly up-down
anti-correlation between the winning branches of successive events
in Fig. \ref{Multiple} is reminiscent of the spatial ordering
observed in the microbranching instability (see Fig. 4c in
\cite{96SF}). This similarity, in conjunction with the resemblance
discussed in relation with Fig. \ref{BranchLength}, suggests that
our 2-dimensional theory might have captured some of the features of
the microbranching instability, although it cannot be directly
applied in that case (see Sect. \ref{micro}).

%%%%%%% FIGURE 8 %%%%%%%%%%%%%%%%%%
%%%%%%%%%%%%%%%%%%%%%%%%%%%%%%%%%%

The crack pattern shown in Fig. \ref{Multiple} reveals a definite
time scale $\tau$ which separates successive branching events. We
can understand this time scale with the following argument: at
branching, the length $L$ of the crack is much larger than any other
length scale like $\ell$ or $\delta\ell$. Accordingly, the stress
intensity factors that determine the velocities are very weak
functions of $\ell$. On the other hand, we expect the pattern of
events to be of a linear size of the order of $\ell$. For a fixed
$\delta\ell$ we thus expect $\tau$ to be of the order of $\ell$.
Therefore, for a general perturbation $\delta\ell$ we expect
\begin{equation}
\tau\sim \ell~ f\left(\case{\delta\ell}{\ell}\right) \ . \label{tau}
\end{equation}
In passing we note the relevance of the prediction (\ref{tau}) to
experiments. In real materials there are always inhomogeneities and
asperities that introduce the perturbations that were modelled here
by $\delta\ell$. Denoting the density of such asperities by $\rho$,
we then estimate the length $\ell$ before meeting the first asperity
by ${\rho}^{-0.5}$. The size of the asperity will determine
$\delta\ell$. It appears worthwhile then to test the relation
(\ref{tau}) in light of this identification.

\subsection{Symmetric Branching}

As discussed above, at higher branching velocities, symmetric
branches can coexist for a longer time, and it is interesting to
determine the typical profiles of such symmetric branches. Consider
an experiment in which a crack of length $L$ in a long strip of
width $W$ bifurcates and two symmetric branches of length $\ell$
emerge. As the branches start propagating with $K_{_{II}} \approx
0$, they cannot change their direction as long as $\ell \ll L,W$. On
the other hand, when $\ell$ grows to the order of the smaller
between $L$ and $W$, the branches will curve. In our theory there is
only one length $L$ (the system is infinite) and we study the
curving of branches of length comparable to $L$.

\begin{figure}
\centering \epsfig{width=.45\textwidth,file=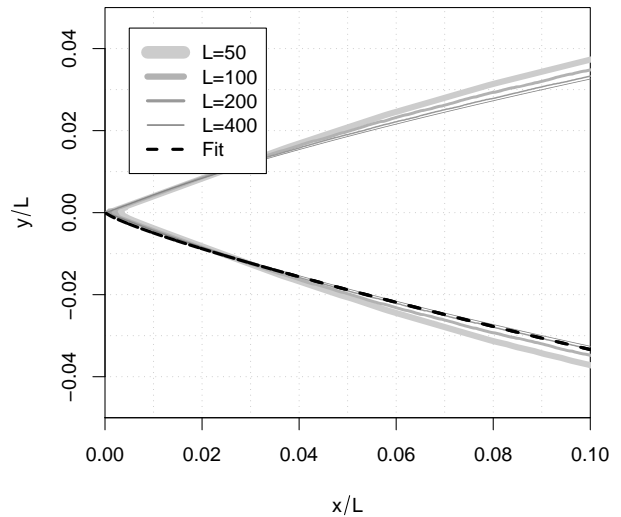}
\caption{Symmetric branching dynamics for different branching length
$L$ with $V_b=0.5$. The plot shows the various crack patterns in
rescaled coordinates. A power law $y \sim x^{\zeta}$ with $\zeta =
0.8$ was added as a guide for the eye.}\label{SymBranch}
\end{figure}

Figure \ref{SymBranch} shows several branching scenarios for
different branching length $L$. It is tempting, after the example of
\cite{95SGF}, to fit power laws to these profiles. The different
profiles can be approximated in the limited range that we consider
by a power law $y \sim x^{\zeta}$ with $0.7 < \zeta < 0.8$. The fit
in the figure corresponds to $\zeta = 0.8$ and was added as a guide
for the eye.
%%%%%%% FIGURE 9 %%%%%%%%%%%%%%%%%%
%%%%%%%%%%%%%%%%%%%%%%%%%%%%%%%%%%%
It should be immediately said however that the profiles are neither
universal nor true power laws. They represent a transient behavior
between two straight lines. Initially the branches start off with an
angle $\lambda\pi =0.13\pi$. Finally there is an asymptotic fixed angle
that depends on the geometry of the system. In an infinite medium
this final angle satisfies $0 < \lambda < 0.13$, while in a strip of
finite width $\lambda=0$, i.e. the branches propagate eventually
parallel to the boundaries of the strip.

\begin{figure}
\centering \epsfig{width=.45\textwidth,file=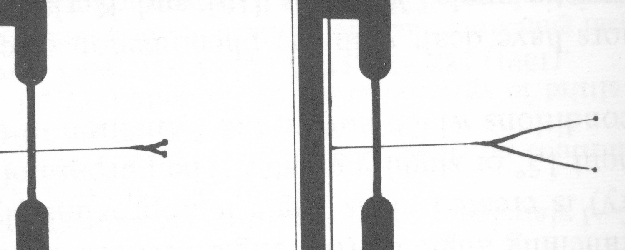}
\caption{Two shadow photographs of a symmetric branching event in a
$300\times 100\times 9$ mm glass plate \cite{73Kal}. The left panel
shows the early stages of the branching process where the branches
are almost straight, while the right panel shows the developed
curved branching configuration.} \label{Kalthoff2}
\end{figure}

As an example of the relevance of our calculation to actual crack
branching events, Fig. \ref{Kalthoff2} shows two shadow photographs
of different stages of a symmetric branching event in a glass plate
\cite{73Kal}. The left panel shows the onset of branching with
straight branches as long as $\ell$ is smaller than $L$ and $W$. The
right panel shows the developed branching configuration. The strong
similarity with Fig. \ref{SymBranch} is obvious.
%%%%%%% FIGURE 10 %%%%%%%%%%%%%%%%%%
%%%%%%%%%%%%%%%%%%%%%%%%%%%%%%%%%%%
\section{How about Microbranching?}
\label{micro} With the relative success of the model proposed here
in reproducing the dynamics and geometry of macrobranches it is of
course tempting to see whether also the geometry of microbranches
can be gleaned from the present 2-dimensional theory. To this aim we
have attempted to follow the dynamics of a side branch in an
asymmetric configuration in which the main branch is forced to
emerge in the forward direction, see the inset of Fig. \ref{asymm}.
We selected $\lambda = 0.2$ such that $K_{_{II}}$ of the side branch
is approximately zero. This initial configuration was introduced as
a constraint on the system to reflect the local symmetry breaking
observed in experiments \cite{95SGF}. Generally, this asymmetry
results in a velocity difference between the side branch and the
main branch.

%%%%%%% FIGURE 11 %%%%%%%%%%%%%%%%%%
\begin{figure}
\centering \epsfig{width=.55\textwidth,file=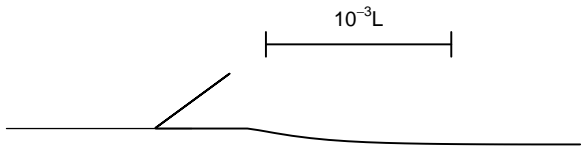}
\caption{A typical resulting configuration for asymmetric branching. The side
branch is almost immediately arrested, while the main branch
temporarily deflected from its straight path. The net effect of such
an event is a strong fluctuation in the velocity of the main branch.
The bar shows the scale of the process relative to the initial crack
length $L$.}\label{AsymCrack2}
\end{figure}
%%%%%%%%%%%%%%%%%%%%%%%%%%%%%%%%%%%

Figure \ref{AsymCrack2} shows a typical resulting configuration in
which the velocity of the main branch fluctuates in a way similar to
the experimentally observed velocity fluctuations, while the side
branch was almost immediately arrested. Since we were mainly
interested in the dynamics of the side branch, i.e. its trajectory
and life time, we concluded that a model that treats the main
macroscopic branch and the microscopic side branch on equal footings
is doomed to fail. The meaning of this result is that the energy
flux into the microscopic side-branch is dramatically {\em
underestimated} in a 2-dimensional model. We tried to increase
artificially the energy release rate $G$ into the near tip region of
the side-branch by a constant factor, such as to increase its initial
velocity. Nevertheless, the velocity of the side branch dropped
immediately to zero, reflecting the huge screening effect of the
main branch. Therefore, although some of our results in Sect.
\ref{moredym} show similarities with various features of the
microbranching instability, we propose that the phenomenon of
microbranching is essentially 3-dimensional and cannot be modelled
directly by a 2-dimensional theory.

%%%%%%%%%%%%%%%%%%%%%%%%%%%%%%%
\section{Summary and Conclusions}
\label{summary}

In summary, we have introduced dynamical equations of motion for
crack tips which depend only on the stress intensity factors at the
tips. Generally speaking, the calculation of these objects is
daunting. By adopting an approximate separable form for the dynamic
stress intensity factors in terms of their static counterparts and
universal velocity dependent functions we achieved a huge
simplification that results in tractable dynamics. Instead of
complicated field equations we can reduce the theory to ordinary
differential equations for the crack tips. Complex events like crack
bifurcations, branch competition, branch arrest and successive
bifurcations are studied in detail and compared with experiments.
The good news is that the comparison with experiments is very
encouraging. The bad news is that the quality of the approximation
cannot be easily assessed from first principles. One expects that
for low velocities and small distances between the multiple crack
tips the approximation should be quite good. What is the range of
validity can at this point be gleaned only from comparison with
experiments, and these are relatively old and not detailed enough,
maybe giving the false impression of a good agreement. It thus
seems very worthwhile to conduct {\em new} experiments in light of
this theory in order to test it in some detail. The gained
transparency and simplicity of the theory seems a very good
motivation for such an undertaking.

\acknowledgments This work had been supported in part by the
European Commission under a TMR grant, The Israel Science Foundation
administered by the Israel Academy, and the Minerva Foundation,
Munich Germany.

\begin{appendix}
\label{appendix}
\section{Static Mode III asymmetric branching}

The aim of this appendix is to derive the static stress intensity
factors at the tips of an asymmetric branched mode III
configuration. Since the mode III problem is described by Laplace
equation for the displacement field $u_z(x,y)$
\begin{equation}
\label{Laplace}
  \nabla^2 u_z(x,y)=0 \ ,
\end{equation}
 the solution
of this problem is readily given if the conformal map,
$z=\Phi(\omega)$, from the exterior of the unit circle to the
exterior of a branched crack configuration is known. The general
formalism for obtaining such a map is known in the literature for a
long time \cite{69And}. Here we adapt the general formalism to the
problem at hand and solve it. Consider the following map
\begin{equation}
\label{map}
\Phi(\omega)=A\omega^{-1}(w-e^{i\alpha_1})^{\lambda_1}(w-e^{i\alpha_2})^{\lambda_2}
 (w-e^{i\alpha_3})^{\lambda_3} \ ,
\end{equation}
where $A$ is a real constant, $0<\alpha_1<\alpha_2<\alpha_3<2\pi$
and $\lambda_3=2-\lambda_1-\lambda_2$. The points
$\{e^{i\alpha_k}\}$ are mapped to the origin and therefore are
branch points. For $\alpha_{k-1}<\theta<\alpha_{k}$ the phase of
$\Phi(\omega)$ is fixed by the crack branch angle and a local
maximum of $|\Phi(\omega)|$ is obtained at $e^{i\beta_k}$. The
parameters of the map, i.e. $A,~ \{\alpha_k\}$ and $\{\beta_k\}$, can
be calculated by demanding that for $\omega=1$ $\arg(z)=0$ and by
the conditions $\Phi'(e^{i\beta_k})=0$,
$|\Phi(e^{i\beta_k})|=\ell_k$. Here $\{\ell_k\}$ are the lengths of
the crack branches. These three conditions can be translated into
the following set of equations
\begin{eqnarray} \label{equations}
 &\sum\limits_{j=1}^{3}&\alpha_j \lambda_j=2\pi \nonumber\\
 &\sum\limits_{j=1}^{3}& \lambda_j
\cot{\left(\frac{\alpha_j-\beta_k}{2}\right)}=0 \nonumber\\
 4A &\prod\limits_{j=1}^{3}&
\left|\sin{\left(\frac{\beta_k-\alpha_j}{2}\right)}\right|^{\lambda_j}=\ell_k
\ ,
\end{eqnarray}
which can be solved numerically. Having the conformal map for the
required configuration at hand we are mainly interested in the
stress intensity factors for the various crack tips. The solution
for Eq. (\ref{Laplace}) is given by
\begin{equation}
 u_z(x,y)=\frac{1}{2\mu}\left[\varphi(z)+\overline{\varphi(z)}\right]
\
,
\end{equation}
where $\varphi(z)$ is an analytic function. The stress components in
polar coordinates are given by
\begin{equation}
\label{stress}
  \sigma_{rz}(r,\theta)-i\sigma_{\theta
  z}(r,\theta)=e^{i \theta} \varphi'(r e^{i\theta})=e^{i \theta}
   \frac{\tilde \varphi '(\omega)}{\Phi'(\omega)} \ .
\end{equation}
Here the tilde denotes the usual transplantation. In the near
vicinity of a given crack tip $z_k$ we use the following expansion
\begin{eqnarray}
 \sigma_{rz}(r,\theta)&=&\frac{K_{_{\rm III}}}{\sqrt{2\pi
 r}}\sin{(\theta/2)} + {\cal O}(\sqrt{r}) \nonumber\\
 \sigma_{\theta z}(r,\theta)&=&\frac{K_{_{\rm III}}}{\sqrt{2\pi
 r}}\cos{(\theta/2)} + {\cal O}(\sqrt{r}) \ .
\end{eqnarray}
These two equations can be rewritten as
\begin{equation}
\label{near tip}
 \sigma_{rz}(r,\theta)-i
 \sigma_{\theta z}(r,\theta)= \frac{-i e^{i \theta/2} K_{_{\rm III}}}
{\sqrt{2\pi
 r}} + {\cal O}(\sqrt{r}) \ .
\end{equation}
On the other hand, near the crack tips we can expand the conformal
map to obtain
\begin{eqnarray}
z-z_k&=&\Phi(\omega)-\Phi(\omega_k)\simeq
\frac{1}{2}\Phi''(\omega_k)(\omega-\omega_k)^2 \nonumber\\
\Phi'(\omega) &\simeq& \Phi''(\omega_k)(\omega-\omega_k) =
\sqrt{2\Phi''(\omega_k) (z-z_k)} \nonumber\\ \ .
\end{eqnarray}
The last expression, in the light of Eq. (\ref{stress}), shows the
explicit relation between the square root singularity of the stress
field near the crack tip and the derivative of the conformal map.
Let us denote $z-z_k=re^{i (\delta_k \pi+\theta)}$ with
$\delta_k=\sum_{j=1}^{k-1}\lambda_j$ and consider the direction
tangent to the crack tip, i.e. $\theta=0$. By comparing Eq.
(\ref{stress}) with Eq. (\ref{near tip}) we obtain
\begin{equation}
\label{SIF}
K_{_{\rm III}}=i \tilde \varphi '(\omega_k)
\sqrt{\frac{\pi}{\Phi''(\omega_k)
e^{i \delta_k \pi}}} \ .
\end{equation}
With this result at hand we can calculate the stress intensity
factor at each tip since the solution in the $\omega$-plane is known
to be
\begin{equation}
\label{solution}
 \tilde \varphi(\omega) = -i \sigma_{yz}^{\infty} A
\left[\omega-\omega^{-1}
 \right] \ ,
\end{equation}
where $\sigma_{yz}^{\infty}$ is the applied stress at infinity. Now
we are in a position to analyze infinitesimal asymmetric branched
configurations. We choose $\ell_1=L$, $\ell_2=\ell_3=\ell$,
$\lambda_1=1-\lambda$ and $\lambda_2=\lambda$ with $\ell/L=0.5\times
10^{-3}$.
The resulting stress intensity factors are presented in Fig.
\ref{AsymmSIFModeIII} in the text.
\end{appendix}

\end{document}